\documentclass[letterpaper,twocolumn,10pt]{article}
\usepackage{usenix-2020-09}

\usepackage{tikz}
\usepackage{amsmath}
\usepackage{multirow}
\usepackage{multicol}
\usepackage{xspace}
\usepackage{wrapfig}
\usepackage{ifthen}
\usepackage{hyperref}
\usepackage{amssymb}
\usepackage{pifont}
\usepackage{cuted}
\usepackage{caption}
\usetikzlibrary{positioning,arrows.meta}

\newboolean{showcomments}
\setboolean{showcomments}{true}

\newcommand{\smartparagraph}[1]{\noindent{\bf #1}\ }
\newcommand{\leakedresult}[1]{\underline{\textcolor{red}{#1}}}
\newcommand{\system}{netFound\xspace}

\begin{document}

\title{\system: From Diagnostics-Informed Design to Production-Ready Network Foundation Model}

\title{\system: Principled Design for Network Foundation Models}


\author{
{\rm Sylee (Roman) Beltiukov\thanks{Both authors contributed equally.}}\\
UC Santa Barbara
\and
{\rm Satyandra Guthula\footnotemark[1]}\\
UC Santa Barbara
\and
{\rm Haarika Manda}\\
UC Santa Barbara
\and
{\rm Jaber Daneshamooz}\\
UC Santa Barbara
\and
{\rm Wenbo Guo}\\
UC Santa Barbara
\and
{\rm Walter Willinger}\\
Northwestern University
\and
{\rm Arpit Gupta}\\
UC Santa Barbara
\and
{\rm Inder Monga}\\
ESNet
} 

\maketitle

\begin{abstract}

Network foundation models promise reusable representations for diverse traffic analysis tasks, but recent diagnostic works have revealed fundamental problems: models exploit dataset shortcuts rather than learning genuine traffic patterns, produce collapsed embedding spaces, and fail to capture the exogenous network conditions that shape real-world behavior. We translate these diagnostic insights into four concrete design principles: protocol-aware tokenization, operational context embedding, burst-flow hierarchical attention, and privacy-by-construction input design, and build \textit{\system}, a network foundation model whose architecture is motivated by this failure analysis. We pretrain \system on a billion-token-scale corpus over 5000 GPU hours, and demonstrate that it produces high-quality representations with lower anisotropy, significantly higher alignment with domain-expert features, and an $F_1$ of 0.95 on exogenous context discrimination where existing state-of-the-art models score below 0.62, while \textit{preserving privacy by excluding payload and IP addresses}. \system demonstrates significant improvements in frozen-encoder evaluation, showing that pretrained embeddings themselves carry useful structure, and remains the top performer across all benchmarks in end-to-end fine-tuned settings. We release full open-source code, weights for three model sizes on HuggingFace, a containerized pipeline from raw PCAPs to downstream inference, and the \textit{full 4.2 billion flows pretraining dataset} to facilitate reproducibility and further research.

\end{abstract}

\section{Introduction}



\smartparagraph{Deployable network intelligence depends on trustworthy traffic representations.}
Self-driving networks, capable of monitoring, diagnosing, and optimizing themselves with minimal human intervention, have been a long-standing aspiration of the networking community~\cite{arxiv-self-driving,nsf-self-driving,mestres2017knowledge}. Recent advances in large language models have brought this vision closer: LLMs can parse network configurations, generate management code~\cite{mani2023hotnets}, and reason about network intent~\cite{netllm2024sigcomm}. However, application of text-based LLMs to raw traffic data is limited and network operators increasingly need models that can turn packet traces into reusable representations for tasks such as traffic classification, anomaly detection, diagnosis, or even agentic-based decision making. 
With over 95\% of web traffic encrypted under TLS~\cite{curtains} and due to privacy considerations, in production these representations must be learned from packet headers, timing, and flow behavior rather than payload, which limits the information available for analysis. All these requirements call for more trustworthy, reusable traffic representations that can be learned at scale and deployed in real-world workflows beyond pure benchmark performance.

\begin{figure}[t]
\centering
\resizebox{0.8\linewidth}{!}{%
    \begin{tikzpicture}[
    font=\footnotesize,
    >=Latex,
    node distance=2mm and 15mm,
    box/.style={
        draw,
        rounded corners,
        align=center,
        fill=red!5,
        text width=3cm
    },
    boxL/.style={
        box,
        fill=blue!5,
    },
    link/.style={-, semithick}
]

\node[boxL] (f1) {Shortcut exploitation};
\node[boxL, below=of f1] (f2) {Collapsed embeddings};
\node[boxL, below=of f2] (f3) {Missing\\exogenous context};

\node[box, right=of f1, yshift=5mm] (p1) {Protocol-aware tokenization};
\node[box, below=of p1] (p2) {Operational\\context embedding};
\node[box, below=of p2] (p3) {Burst-flow\\hierarchical attention};
\node[box, below=of p3] (p4) {Privacy-by-construction input design};

\draw[link] (f1.east) -- (p1.west);
\draw[link] (f1.east) -- (p4.west);

\draw[link] (f2.east) -- (p1.west);
\draw[link] (f2.east) -- (p2.west);
\draw[link] (f2.east) -- (p3.west);

\draw[link] (f3.east) -- (p2.west);
\draw[link] (f3.east) -- (p3.west);
\draw[link] (f3.east) -- (p4.west);

\end{tikzpicture}
}
\caption{Design principles (right) motivated by diagnostic findings (left).}
\label{fig:findings_principles_map}
\end{figure}

\smartparagraph{Existing network foundation models fail to yield reusable traffic representations.}
Network foundation models (NFMs) promise reusable traffic encoders through large-scale self-supervised pretraining~\cite{etbert,yatc,flow-mae,netmamba,trafficformer}. Yet recent diagnostic works show that many current NFMs function at best as benchmark solvers and yield at worst non-reusable representations. Trustee~\cite{trustee} showed that network ML systems often exploit dataset shortcuts rather than learning genuine traffic behavior. Pcap-Encoder~\cite{debunking} demonstrated that NFM results can be inflated by leakage pathways such as per-packet train/test splits and timestamp correlations. The Intrinsic Evaluation Framework (IEF)~\cite{demystifying} further showed that existing NFMs often produce anisotropic (collapsed) embeddings, align weakly with domain-expert traffic features, and fail to capture exogenous network conditions that shape observed behavior. Together, these results suggest that current NFMs are not yet a sufficient foundation for privacy-preserving, production-facing network understanding.

\smartparagraph{The missing piece is a network-native, diagnostics-informed design.}
These failures are not incidental. Most existing NFMs inherit tokenization, embedding, and attention mechanisms from natural language processing or vision, even though network traffic has protocol-defined field structure, rich operational metadata, a natural burst-flow hierarchy, and strict privacy constraints. Our thesis is that a deployable network foundation model must be designed around these properties from the start. It should preserve protocol semantics during tokenization, encode metadata alongside header content, model behavior at both burst and flow scales, and restrict identity-bearing inputs by construction.~\autoref{sec:design-principles} formalizes these requirements as four design principles that guide the rest of the paper.

\smartparagraph{\system is our answer: a production-ready, privacy-preserving network foundation encoder.}
We present \textit{\system}, an encoder-only, network-native foundation model built around this design thesis. \system applies protocol-aware tokenization, injects operational context into token representations, uses burst-flow hierarchical attention to capture multi-scale traffic structure, and operates on packet headers without payload or endpoint IP identities. Beyond the model architecture, \system is built as a practical system: we pretrain three model sizes on a billion-token-scale corpus, release full code, pretraining dataset, weights, and provide a containerized end-to-end pipeline from raw PCAPs to downstream inference so that the model can be reproduced, extended, and deployed rather than treated as a one-off artifact.

\smartparagraph{This design yields stronger reusable representations in both intrinsic and downstream evaluation.}
Using the same diagnostic lenses that revealed the limitations of prior work, we show that \system produces substantially better representations. \system reaches an $F_1$ of 0.95 on exogenous context discrimination, compared to 0.62 for the best prior state-of-the-art baseline, and it is the strongest model by a wide margin in frozen-encoder evaluation, demonstrating that the pretrained embeddings themselves carry useful structure before task-specific adaptation. At the same time, on downstream benchmarks, \system remains top-performing across five tasks in both frozen and end-to-end fine-tuned settings. These results show that privacy-by-construction input design, representation quality, and production-readiness can be improved together rather than traded off against one another.

\subsubsection*{Contributions.}
\begin{enumerate}
    \item \textbf{From diagnosis to deployable design.} We show how recent diagnostic findings on shortcut exploitation, weak embedding structure, and poor context sensitivity~\cite{trustee,debunking,demystifying} translate into concrete requirements for a deployable network foundation model, and we formalize these requirements as four design principles (\S\ref{sec:design-principles}).

    \item \textbf{A production-ready, privacy-preserving network foundation encoder.} We build \system, a network-native encoder that implements these principles through protocol-aware tokenization, operational context embedding, burst-flow hierarchical attention, and privacy-by-construction input selection, and we pair it with an end-to-end reproducible system for preprocessing, pretraining, fine-tuning, and deployment (\S\ref{sec:tech_design}, \S\ref{sec:implementation}). We fully release open-source implementation under permissive license, including pretrained weights for three different model sizes (53M--663M parameters) on HuggingFace, a containerized end-to-end pipeline, native C++ acceleration for preprocessing, and an automated test suite (\S\ref{sec:implementation}), to enable the community to reproduce, extend, and deploy \system rather than treating it as a one-off artifact. We also publish the \textit{full pretraining dataset}, containing \textit{4.2 billion} anonymized network flows in the Apache Arrow format, suitable for \system pretraining for full reproducibility of the results. 

    \item \textbf{Evidence of stronger reusable representations.} We demonstrate, through intrinsic evaluation and downstream benchmarks, that \system yields higher-quality reusable traffic representations than prior NFMs. These representations especially shine in the better discrimination of exogenous network context compared to six existing models (including Pcap-Encoder) and in frozen-encoder downstream settings, while remaining top-performing after full end-to-end fine-tuning across five downstream tasks (\S\ref{sec:eval-intrinsic}, \S\ref{sec:eval-fine}).
\end{enumerate}

\section{Background and Problem Scope}

This section reviews representation learning in networking and identifies the problems in current network foundation model designs that motivate our approach.


\subsection{Representation Learning in Networking}

The core premise of representation learning is that raw data (packet captures) can be transformed into compact vector embeddings that capture the underlying structure relevant to downstream tasks. A good network representation should satisfy four requirements derived from the failure modes identified by recent diagnostic work~\cite{demystifying,debunking}: (1)~\textit{semantic preservation}, where protocol field boundaries and their meanings survive tokenization; (2)~\textit{operational grounding}, where the representation encodes the operational metadata (timing, burst statistics, directionality) that domain experts rely on;  (3)~\textit{contextual depth}, where the representation captures not only what is in the packet headers but also the exogenous conditions, such as congestion control, queue management, or cross-traffic, that shape traffic behavior; and (4)~\textit{leakage resilience}, where the representation avoids dependence on fields that leak identity or split-specific information, such as IP addresses, checksums, payload bytes, or other fields that a model could memorize to artificially inflate generalization.

Existing approaches to network representation learning fall into two categories. Task-specific methods extract hand-crafted features and train a supervised model per task~\cite{mirskyKitsuneEnsembleAutoencoders2018,ahmadNetworkIntrusionDetection2021}; they can be effective but do not generalize across tasks and require labeled data. Foundation models instead pretrain on large unlabeled corpora and produce reusable representations that can be adapted to many downstream tasks via fine-tuning. The promise is clear, but as we show next, both the models and their representations are seriously flawed.

\subsection{Diagnosis: Why Existing Models Fall Short}
\label{sec:diagnosis}

Inspired by the success of foundation models in other domains, multiple network foundation models have been developed in recent years~\cite{pert,etbert,yatc,flow-mae,mtsecurity,trafficgpt,lens,debunking}. These solutions vary in tokenization, embedding, and pretraining strategies, but share a common root cause: most treat network data as natural language or images and employ transformer architectures developed for these domains. ET-BERT treats network data as natural language using BERT with domain-specific pretraining tasks. YaTC~\cite{yatc} and Flow-MAE~\cite{flow-mae} treat data as images using Vision Transformer~\cite{visiontransformer} and Masked Autoencoders~\cite{maskedautoencoder}. Pcap-Encoder~\cite{debunking} adopts a T5-based architecture with a corrected per-flow evaluation protocol, but uses simple packet hexadecimal representation.
\begin{table}[t]
    \centering
    \caption{Design choices across network foundation models. Existing models import NLP/vision defaults; \system makes domain-specific choices for each dimension.}
    \resizebox{\columnwidth}{!}{
        \begin{tabular}{l|llll}
        \textbf{Model} & \textbf{Tokenization} & \textbf{Metadata} & \textbf{Attention} & \textbf{Input} \\
        \hline
        ET-BERT~\cite{etbert}         & 2B BPE (payload)   & ---  & flat        & payload bytes \\
        PERT~\cite{pert}              & 2B chunks (payload) & --- & flat        & payload bytes \\
        LENS~\cite{lens}              & fixed-byte (payload) & --- & flat       & payload bytes \\
        TrafficGPT~\cite{trafficgpt}  & byte-level (payload) & --- & flat       & payload bytes \\
        YaTC~\cite{yatc}              & fixed-byte (hdr+pay) & --- & pkt$\to$flow & hdr+payload \\
        Flow-MAE~\cite{flow-mae}      & fixed-byte (hdr+pay) & --- & pkt$\to$flow & hdr+payload \\
        Pcap-Encoder~\cite{debunking} & fixed-byte (hdr+pay) & --- & flat       & hdr+payload \\
        \hline
        \textbf{\system}             & \textbf{protocol-field} & \textbf{IAT, burst, dir.} & \textbf{burst$\to$flow} & \textbf{anon. headers} \\
        \end{tabular}
    }
    \label{table:models}
\end{table}
\autoref{table:models} summarizes how each model handles the four design dimensions. Existing models default to generic NLP/vision choices instead of making domain-specific decisions for each. Two recent diagnostic studies reveal the consequences of these design gaps.

\smartparagraph{Intrinsic representation failures (IEF).}
The Intrinsic Evaluation Framework~\cite{demystifying} introduced a task-agnostic methodology for probing NFM representations through three lenses. \emph{Embedding geometry}: mean cosine similarity between flow embeddings reveals that existing models produce highly anisotropic representations, meaning diverse flows are mapped to nearly identical vectors, making it impossible for consequent layers to separate. \emph{Metric alignment}: The centered kernel alignment (CKA) index that measures similarity between model embeddings and CICFlowMeter domain-expert features~\cite{lashkariCharacterizationTorTraffic2017} is near zero, indicating that representations fail to encode the operational statistics that network engineers use. \emph{Exogenous context discrimination}: linear probes trained on flow embeddings achieve low $F_1 < 0.62$ for discriminating congestion control algorithms, queue management policies, and cross-traffic patterns. These conditions significantly shape traffic behavior but are usually invisible in packet headers, making it a challenging task for ML models to work with.

\smartparagraph{Evaluation pitfalls and shortcut exploitation (Pcap-Encoder).}
Pcap-Encoder~\cite{debunking} and Trustee~\cite{trustee} demonstrated that models trained on network datasets frequently rely on shortcuts and spurious correlations rather than learning meaningful traffic patterns. Among common examples, per-packet splitting allows models to use TCP sequence numbers as flow identifiers, and timestamp correlations in simultaneously-collected datasets often leak application identity.
\begin{table}[t]
    \centering
    \caption{$F_1$ score of models on original versions of datasets (with shortcuts) and fixed (without shortcuts). Performance drop signals vulnerability to shortcuts.}
    \resizebox{\columnwidth}{!}{
        \begin{tabular}{l|cc|cc}
        & \multicolumn{2}{c|}{\textbf{Crossmarket ($Acc@10$)}} & \multicolumn{2}{c}{\textbf{CIC-IDS (Heartbleed)}} \\
        & Original & Fixed & Original & Fixed \\
        \hline
        ET-BERT
            & $99.82 \pm 0.03$ & $32.83 \pm 0.17$ 
            & $99.99 \pm 0.01$ & $0.0 \pm 0.0$ \\
        YaTC
            & $99.69 \pm 0.03$ & $63.48 \pm 0.12$ 
            & $99.99 \pm 0.01$ & $0.01 \pm 0.01$ \\
        Pcap-Encoder
            & $97.18 \pm 0.21$ & $86.99 \pm 0.32$ 
            & $99.99 \pm 0.00$ & $0.0 \pm 0.0$ \\
        \system-large
            & $65.38 \pm 0.81$ & $65.38 \pm 0.81$ 
            & $99.98 \pm 0.01$ & $99.98 \pm 0.01$ \\
        \end{tabular}
    }
    \label{tab:resilience}
\end{table}
%
%
To demonstrate the resulting problems in generalizability, we find and fix shortcuts for two commonly used downstream datasets: Crossmarket~\cite{crossmarkets} and CIC-IDS-2017~\cite{cic-ids-2017}, by removing the TCP Options field that contains class-unique timestamp in Crossmarkets and fixing the Heartbleed shortcut by splitting the flow between attacks as proposed by Trustee~\cite{trustee} for CIC-IDS-2017.
As shown in \autoref{tab:resilience}, existing models achieve near-perfect performance on original datasets but performance drops dramatically on fixed versions, confirming heavy reliance on spurious features. Pcap-Encoder~\cite{debunking} recognized this problem and proposed several corrections in the evaluation protocol, being able to resolve some of the well-known shortcuts, such as Crossmarket, but not all of them. However, while Pcap-Encoder addresses the evaluation methodology, it retains a protocol-agnostic T5-based architecture: it does not use protocol-aware tokenization, operational metadata, or network-suitable attention designs. This means Pcap-Encoder corrects \emph{how} models are evaluated but does not address \emph{why} they produce poor representations in the first place.
\label{sec:background-shortcuts}

\subsection{From Diagnosis to Design Principles}
\label{sec:design-principles}

The diagnostic findings from preliminary works form a coherent picture of what network foundation models must do differently. We synthesize these findings into four design principles that guide our overall design and are illustrated in \autoref{fig:findings_principles_map}. Crucially, the mapping is many-to-many: each principle addresses multiple diagnostic findings, and each finding motivates multiple principles. In effect, this interconnected structure accentuates why \system's design is principled rather than ad-hoc.

\smartparagraph{Principle 1: Tokenization must respect protocol structure.}
IEF's finding that embeddings have near-zero CKA alignment with domain-expert features, combined with Pcap-Encoder's finding that fixed-sized byte chunks strategy corrupts protocol field semantics, jointly motivate a tokenization strategy that preserves field boundaries. CICFlowMeter features are computed from specific protocol fields (flags, IP TTL, etc.); if tokenization destroys these fields, the model can never learn features that align with them. Furthermore, protocol-aware tokenization enables selective exclusion of fields known to leak identity, providing a structural defense against shortcut exploitation. This motivates \emph{protocol-aware tokenization} (\S\ref{sec:tokenization}).

\smartparagraph{Principle 2: Representations must encode operational metadata.}
IEF showed that existing model embeddings have near-zero alignment with the operational features that domain experts rely on (CKA $< 0.07$), and that models cannot discriminate exogenous network conditions ($F_1 < 0.62$). Both failures share a root cause: models ingest only raw bytes and have no explicit representation of the timing, burst statistics, and directional metadata through which exogenous conditions manifest. IEF demonstrates that models struggle with these features the most, resulting in overall reduced quality of the embeddings. Injecting this metadata also provides additional axes of variation that can help spread representations in the latent space, mitigating the anisotropy problem.
This motivates \emph{operational context embedding} (\S\ref{sec:metadata}).

\smartparagraph{Principle 3: Attention must capture multi-scale temporal structure.}
IEF's context discrimination results reveal that existing models cannot capture exogenous network conditions that manifest at different temporal scales: congestion control affects burst-level inter-arrival times (milliseconds), while cross-traffic patterns shape flow-level behavioral trends (seconds). Ignoring these temporal patterns leads to poor representation capabilities, and addition of a simple flat attention over truncated sequences cannot model both scales simultaneously. YaTC and Flow-MAE attempt hierarchy but without parameter sharing across granularities, limiting cross-level reasoning.
This motivates \emph{burst-flow hierarchical attention} with shared parameters and CLS-token propagation (\S\ref{sec:hierarchy}).

\smartparagraph{Principle 4: Input design must prevent shortcut exploitation and privacy issues by construction.}
Pcap-Encoder and Trustee showed that existing models exploit TCP sequence numbers, timestamp correlations, and IP-based identifiers as shortcuts, achieving near-perfect accuracy that collapses under proper evaluation (\autoref{tab:resilience}). In addition, to protect user privacy, we require that a production-ready model is unable to memorize any potentially sensitive information that can be gleaned from either payload data or certain header fields.
Rather than relying on post-hoc regularization, we adopt a prevention-by-design approach: if the model never sees the leaky features or sensitive information, it cannot exploit them.
This motivates \emph{privacy-by-construction input design} (\S\ref{sec:privacy}).

\section{\system Architecture}
\label{sec:tech_design}



The four principles derived in~\autoref{sec:design-principles} are not independent and form an interconnected set that constrains the design space but does not uniquely determine an architecture. In this section we describe how each principle is realized in \system, and for each design choice we discuss the alternatives we considered and the reasoning that guided our decisions.

\subsection{Protocol-Aware Tokenization}
\label{sec:tokenization}

\begin{figure}[htbp]
    \centering
    \includegraphics[width=\linewidth]{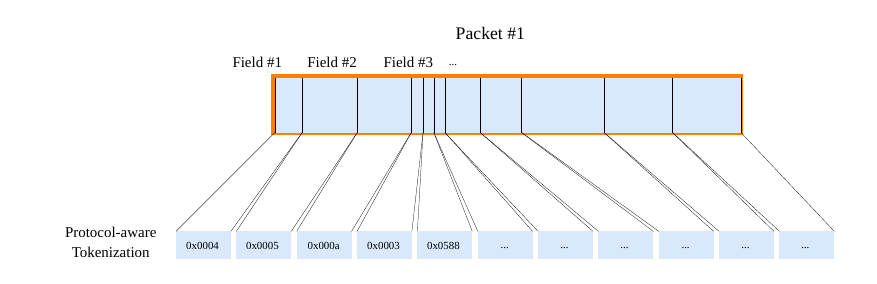}
    \caption{Protocol-aware tokenization preserves packet field semantics by splitting headers along protocol field boundaries.}
    \label{fig:model_overview_tokenizer}
\end{figure}

\smartparagraph{Design space.}
Network foundation models (as shown in \autoref{table:models}) can utilize one of three different strategies for tokenization of raw network data.
(1)~\emph{Fixed-size byte chunks} strategy (which splits a raw byte stream into 2-byte tokens, as in PERT~\cite{pert} and Pcap-Encoder~\cite{debunking}) is simple but protocol-agnostic; e.g., a 2-byte window at the wrong offset splits a 16-bit TCP window-size value across two tokens, forcing the model to learn cross-token reconstruction before it can reason about window behavior.
(2)~\emph{Learned subword tokenization} (e.g., Byte-Pair Encoding, or BPE, creates a translation table for the most popular byte chunks) works by discovering frequently co-occurring byte patterns, but these patterns reflect byte-frequency statistics rather than protocol semantics; BPE cannot guarantee that a token boundary will not bisect a protocol field, and it provides no mechanism for selectively excluding known leaky fields (e.g., IP address) because field identity is not represented in its vocabulary.
(3)~\emph{Protocol-aware tokenization} aligns tokens with protocol field boundaries, ensuring that each token corresponds to exactly one semantically meaningful field. This approach produces correct field splits and associations but requires stricter packet parsing and is computationally more expensive.


\smartparagraph{Preserving field semantics.}
Diagnostic studies~\cite{debunking} have shown that protocol-agnostic fixed-size tokenization is a root cause of semantic corruption in existing network foundation models: token boundaries routinely split protocol fields, destroying the very information the model needs to learn. Because packet headers are structured according to protocols such as TCP, UDP, and IP, we employ strategy (3), \textit{protocol-aware tokenization} strategy, that segments headers based on their protocol-specific fields, as shown in~\autoref{fig:model_overview_tokenizer}. Each token corresponds to exactly one semantically meaningful field (e.g., source port, TTL, TCP flags), allowing variable byte lengths without complicating embedding or model training. This directly addresses Principle~1 from~\autoref{sec:design-principles}: protocol-aware tokenization preserves the semantic units that domain-expert features are computed from, and enables selective exclusion of known leaky fields.

\subsection{Operational Context Embedding}
\label{sec:metadata}

\begin{figure}[htbp]
    \centering
    \includegraphics[width=\linewidth]{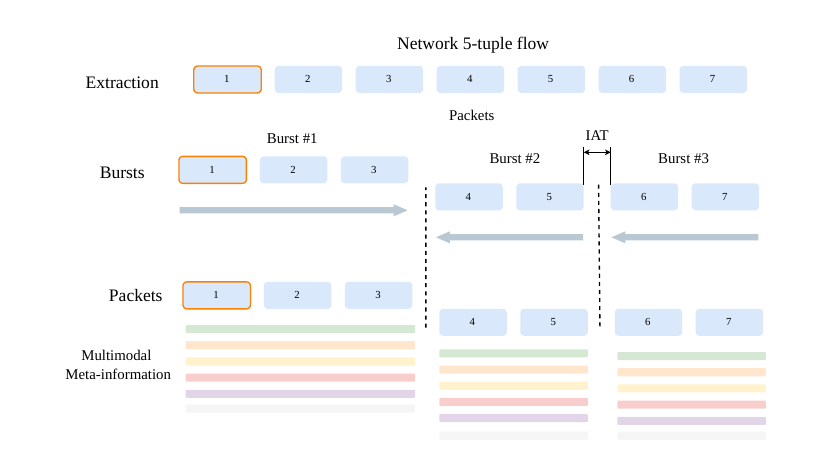}
    \caption{Network-aware data extraction and featurization enriches the model's embeddings with operational metadata.}%
    \label{fig:model_overview_extraction}%
\end{figure}

\smartparagraph{Design space.}
Existing NFMs operate on raw bytes alone, discarding the operational metadata that domain experts rely on and reducing the model's quality. To include this operational metadata, there are several possible approaches.
(1)~\emph{Feature engineering as input} replaces raw bytes with hand-crafted features. By definition, this method achieves high alignment with domain expert-derived features, but the generality of the learned representations suffers: the model can only encode features that were anticipated at design time, and it loses access to raw header content that may serve as a useful but unanticipated signal.
(2)~\emph{Additive embedding} projects metadata into the same space as token embeddings and sums them. While simple, addition is a lossy operation: when the metadata and token embedding vectors occupy overlapping subspaces, the sum conflates the two sources of information and downstream layers cannot recover which contribution came from which modality.
(3)~\emph{Concatenation and compression} preserves both raw bytes and operational data by concatenating them into a wider vector and then projecting back to the model dimension via a learned linear layer. The projection can learn to allocate capacity between modalities, and no information is destroyed before the model has a chance to see both sources.
We adopt approach~(3) because it addresses Principle~2 (grounding representations in operational metadata) while preserving the raw-byte signal that enables the model to discover unexpected features.

\smartparagraph{Grounding representations in domain-expert features.}
Intrinsic evaluation of existing models reveals that their embeddings have near-zero alignment (measured via CKA) with the operational features (timing, burst statistics, directionality) that domain experts rely on for traffic analysis~\cite{demystifying}. Approach (3) allows us to close this alignment gap and encode operational context directly into the token embedding. As shown in~\autoref{fig:model_overview_extraction}, we calculate and embed inter-arrival time information, burst size (number of packets, total number of bytes), direction of the packets in the flow, and relative positions of the packets in the burst and flow. These features are projected through a two-layer MLP with activation into the hidden dimension, then \textit{concatenated} with the word embedding rather than added to it. We additionally embed the transport-layer protocol (e.g., TCP, UDP, ICMP) via a dedicated learned embedding table, broadcast to every token position. A linear projection layer then fuses the three concatenated components (word, metadata, protocol embeddings) back to the model's hidden dimension. This design is similar to the \textit{unified embedding} architecture used in modern multi-modal encoders~\cite{gemini-embedding-2}.

\subsection{Burst-Flow Hierarchical Attention}
\label{sec:hierarchy}

\begin{figure}[htbp]
    \centering
    \includegraphics[width=\linewidth]{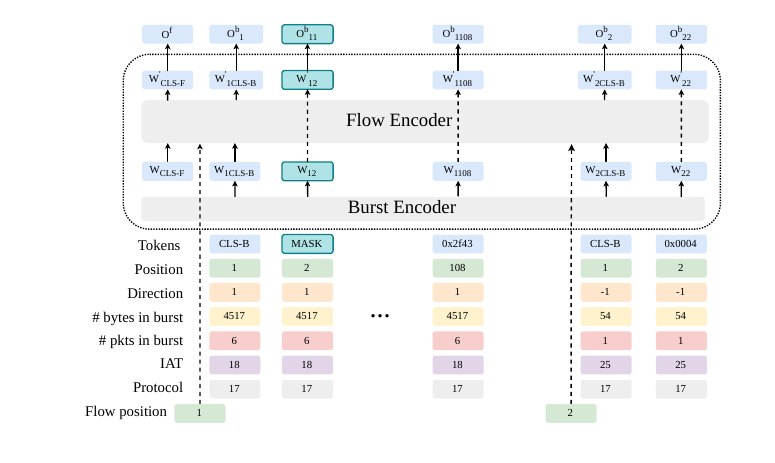}
    \caption{The hierarchical transformer architecture decomposes attention into burst-level and flow-level stages.}%
    \label{fig:model_overview}%
\end{figure}

\smartparagraph{Design space.}
Attention mechanisms give models an opportunity to account for nearby tokens at every step in the calculations. Recent works~\cite{beltagy2020longformer, flashattn} proposed various attention patterns, but the patterns that are suitable for network data fall into four categories: 
(1)~\emph{Flat attention} over the entire token sequence (used by ET-BERT, Pcap-Encoder, etc.) takes all tokens regardless of their position in the burst-flow hierarchy. This attention pattern incurs quadratic cost in sequence length and, more importantly, forces the model to learn the burst-flow structure implicitly from positional embeddings alone. 
(2)~\emph{Windowed/sparse attention} (e.g., Longformer~\cite{beltagy2020longformer}) reduces cost by restricting each token's attention to a fixed window, but the window size remains a hyperparameter and ignores the natural burst boundaries in traffic. A window that is too small misses cross-packet dependencies within a burst; one that is too large conflates adjacent bursts.
(3)~\emph{Hierarchical attention} with burst-level and flow-level stages aligns the attention structure with the data's natural hierarchy: burst encoders capture within-burst packet interactions, flow encoders capture cross-burst behavioral patterns.
(4)~\emph{Using disaggregated models} is an attention mechanism where separate foundation models are trained at the packet, burst, and flow levels and combined at inference. It avoids the architectural complexity of hierarchical attention but prevents cross-level parameter sharing and requires coordinating multiple models at inference time.
We adopt approach~(3) shown on~\autoref{fig:model_overview} because this architecture explicitly models both temporal scales within a single model while enabling cross-level information flow through CLS-token propagation to satisfy Principle~3.

\smartparagraph{Capturing multi-scale dependencies.}
Existing models either flatten entire flows into a single token sequence, losing the natural grouping of packets into bursts, or attempt hierarchy without parameter sharing across granularities~\cite{yatc,flow-mae}. Intrinsic evaluation shows that this prevents models from capturing network conditions that manifest at different time scales~\cite{demystifying}.
Network data exhibits a natural two-level hierarchy: tokens within a burst capture fine-grained packet-field interactions, while bursts within a flow capture coarser temporal and behavioral patterns.
To reflect this behavior, each \system transformer layer is composed of two separate sub-layers, a \textit{burst encoder} and a \textit{flow encoder}, that operate at different granularities.

Given an input sequence of token embeddings, we first reshape the flat token sequence into a burst-structured tensor, grouping tokens by their burst membership.
The burst encoder then applies self-attention independently within each burst, allowing tokens to attend only to other tokens in the same burst.
This step captures local dependencies such as relationships among packet header fields and inter-packet patterns within a burst.
Each burst is prepended with a CLS token whose output representation serves as a holistic summary of that burst.

After the burst encoder, we extract the obtained burst-level CLS representations and feed them into the flow encoder. By applying self-attention across all bursts in a flow, this encoder enables the model to capture cross-burst dependencies such as long-range temporal patterns, flow-level behavioral shifts, and correlations between distant parts of a conversation.
Crucially, individual token representations \textit{skip} the flow encoder via a residual connection and only the CLS tokens are updated by the flow-level attention.
The updated CLS representations are then written back into the full sequence, enriching each burst's summary with flow-level context while preserving the fine-grained token representations.
This hierarchical decomposition offers both computational and modeling advantages.
By restricting burst-level attention to tokens within each burst (typically up to 108 tokens) rather than the entire flow (which can exceed 1,300 tokens), the quadratic cost of self-attention is substantially reduced.
The flow encoder operates only over the burst CLS tokens (i.e., 12), keeping its cost minimal.

\subsection{Privacy-by-Construction Input Design}
\label{sec:privacy}

\smartparagraph{Design space.}
There exist three  popular general input design strategies that define what information a model sees during training and inference.
(1)~\emph{Full packet ingestion} (ET-BERT, YaTC, Pcap-Encoder) feeds the model raw bytes including payload, IP addresses, and all header fields. This strategy maximizes the information available to the model but also maximizes the attack surface for shortcut exploitation: e.g., payload bytes often contain application-level identifiers and IP addresses encode network topology.
(2)~\emph{Post-hoc regularization} retains all features but adds augmentations during training to discourage shortcut use. However, this method is fragile: the model may find new shortcuts not anticipated by the regularizer, and there is no formal guarantee that shortcuts are eliminated.
(3)~\emph{Prevention by design} restricts the input space to features with known behavioral semantics, excluding those that could plausibly leak identity. Doing so sacrifices some information but provides a structural guarantee: the model never sees leaky features, so it cannot exploit them.

Most papers in the existing literature utilize approach~(1) which allows them to demonstrate the highest scores on the fine-tuning datasets, either by shortcut exploitation or memorization of data patterns. However, network data is also known to contain information that can be considered \textit{sensitive} from the users' perspective, such as IP addresses, payload (regardless of encryption), and other info (e.g., domain names in SNI). To protect such information and avoid failures under proper evaluation (as Pcap-Encoder results demonstrated for approaches~(1) and~(2)), we adopt approach~(3) that avoids the use of sensitive or encrypted fields. Prevention by design is the only strategy that addresses Principle~4 with a provable guarantee rather than a statistical one.

\smartparagraph{Preventing shortcut exploitation.}
Concretely, \system operates on packet headers without payload, uses directional information instead of endpoint IP addresses, and applies protocol-aware field selection that can exclude known leaky fields. A modular design allows researchers to include or exclude additional fields (e.g., by using a model's different configurations), enabling them to tailor privacy guarantees to their unique needs.

\smartparagraph{Dynamic padding and bucketing.}
Network flows exhibit high variability in length, both in the number of packets per burst and the number of bursts per flow, posing a challenge for fixed-sized architectures.
We address this through a three-pronged strategy of \textit{truncation}, \textit{dynamic padding}, and \textit{length-bucketed batching}.
First, during tokenization, each burst is truncated to a configurable maximum burst length (i.e., 109 tokens including a CLS token, encoding 6 packets within a single burst), and each flow is truncated to a maximum number of bursts (i.e., 12), bounding the input dimensionality, but still reliably representing the overall flow structure.
Second, rather than padding all sequences to global maximums, our data collator computes per-batch maximums, padding only to those values, avoiding saturating models with empty sequences.
Third, we employ a length-bucketed sampler that groups similarly-sized flows into batches ensuring efficient batch packing.
Together, these mechanisms allow \system to efficiently handle the wide range of sequence lengths inherent in network traffic.

\subsection{Modern Transformer Components}

Following recent advances in efficient transformer design from ModernBERT~\cite{modernbert}, we incorporate several architectural improvements into each transformer layer.
We replace learned absolute position embeddings with Rotary Position Embeddings (RoPE)~\cite{rope}, adopt Flash Attention~\cite{flashattn} and BF16 for accelerated training, and use the ModernBERT MLP design with GeGLU activation~\cite{geglu} and a reduced intermediate size (1.5-2.5$\times$ the hidden dimension).
We also remove bias terms from both the MLP and LayerNorm layers, and apply pre-normalization before the MLP with a residual connection.
Our tokenizer's vocabulary comprises 65,600 entries, of which 65,539 correspond to byte-level token values and model-specific tokens, and the remainder is reserved to allow researchers to use them for specific downstream applications.

\subsection{Pretraining Objectives}

A pretrained encoder must learn representations that are useful across diverse downstream tasks without task-specific supervision. We use a multitask pretraining objective with four complementary losses. Each loss is detailed below and targets a different aspect of traffic structure. The rationale for multiple objectives is that no single self-supervised loss covers all the representational requirements from~\autoref{sec:design-principles}: while the first loss teaches token-level semantics, the remaining three losses teach burst-level and flow-level structure that the first loss alone cannot capture.

\smartparagraph{Masked Language Modeling (MLM).}
Our primary objective follows the masked language modeling paradigm~\cite{devlin2019bert}, adapted for packet header tokens.
During training, 30\% of tokens are selected for masking: 80\% replaced with \texttt{[MASK]}, 10\% with a random token, 10\% unchanged.
The model is trained to reconstruct original tokens using cross-entropy loss.

\smartparagraph{Swapped Burst Detection.}
To encourage flow-level coherence, we introduce a \textit{swapped burst detection} objective.
Here, randomly selected bursts are replaced with ones from a different flow.
A binary classifier on the pooled flow-level representation detects the swap, helping the encoder to learn the burst correlation information.

\smartparagraph{Metadata Prediction.}
We train the model to predict three continuous burst-level metadata features: inter-arrival time, total bytes, and packet count.
For each burst whose metadata is masked (probability 0.3), the model regresses these values from the burst's CLS representation using L1 loss, ensuring that CLS representation learns useful information from flows.

\smartparagraph{Direction Prediction.}
A direction classifier predicts burst direction (inbound vs.\ outbound) on burst CLS representations, further helping the model with flow understanding.

\smartparagraph{Loss Weighting.}
Each loss is independently weighted, and the total pretraining loss is their weighted sum.
This multi-task formulation ensures that \system simultaneously learns fine-grained token-level semantics, burst-level statistical properties, and flow-level coherence.

\section{Implementation}
\label{sec:implementation}



An architecture that satisfies the design principles specified in~\autoref{sec:design-principles} is necessary but not sufficient for a practical network foundation model. From a practical perspective, the engineering decisions matter as much as the architectural choices and concern issues such as how the model is trained, how data flows from raw PCAPs to learned representations, and whether the system can be reproduced and deployed by others. In this section, we describe the key engineering decisions we made  and the reasoning behind each decision.

\subsection{Pretraining Data and Model Variants}

\smartparagraph{Pretraining data.}
We pretrain \system on a dataset collected from a large U.S.\ university campus, comprising multiple weeks of traffic captured at the campus border.
The resulting corpus contains \textbf{four billion} diverse network flows ($\sim$1.2TB after preprocessing), spanning a mix of residential, enterprise, and research traffic.\footnote{We attempted to increase diversity by adding MAWI~\cite{mawi} and CAIDA~\cite{caida} to the mix, but did not observe any significant improvements in terms of intrinsic evaluation or downstream performance and discarded this approach.} The source, scale, and diversity of the pretraining data matter for a number of reasons.

For one, NFMs trained on narrow traffic mixes risk learning dataset-specific or application-specific patterns rather than generalizable traffic representations, and the same shortcut problem that Pcap-Encoder~\cite{debunking} identified at the evaluation level can also occur at the pretraining level. As we show later in \S\ref{sec:eval-fine}, since most of the publicly available models use the same datasets for pretraining and fine-tuning~\cite{etbert, yatc, netmamba, trafficformer}, they benefit from leakage of test data into the pretraining data which, in turn, enables them to achieve excellent performance on the considered fine-tuning tasks.
In addition, the burst-flow hierarchical attention mechanism (\S\ref{sec:hierarchy}) requires sufficient diversity of burst structures to learn meaningful cross-burst patterns rather than memorizing the burst-length distribution of a single network. 
Using diverse network data collected from a production network with tens of thousands of users allows us to expose \system to a wide range of different network conditions and application behaviors, simultaneously avoiding skewing the fine-tuning results and a model's ability to generalize. Limitations of this approach are discussed in~\S\ref{sec:limitations}. 

\smartparagraph{Three model sizes.}
We independently pretrain three model variants: small (four layers and four attention heads, 512-dim, 53M parameters), base (12 layers and 12 attention heads, 768-dim, 225M parameters), and large (24 layers and 16 attention heads, 1024-dim, 663M parameters).
The rationale for multiple sizes is deployment flexibility: a small model may suffice for edge or near-real-time inference where latency is critical, while a large model is appropriate for offline analysis where accuracy matters most.
The efficiency evaluation in \autoref{sec:eval-efficiency} quantifies this trade-off.

\smartparagraph{Multi-stage pretraining.}
We implement multiple pretraining stages, varying loss weights to encourage progressively more complex dependencies: early stages emphasize MLM for token-level semantics, while later stages increase the weight on burst-level and flow-level objectives.
For the entire duration of pretraining, we use AdamW~\cite{adamw} with a Warmup-Stable-Decay learning rate schedule (peak 1e-4), reducing the learning rate on subsequent stages. In total, we utilize 128 NVIDIA A100 GPUs, with approximately 10 billion tokens seen across stages for each model, with flows randomly sampled in a streaming manner from the pretraining dataset.

\subsection{Data Pipeline and Reproducibility}

\smartparagraph{C++ acceleration.}
The data preprocessing pipeline comprises five automated stages: protocol filtering, flow-based PCAP splitting, field extraction, tokenization, and Apache Arrow shard assembly.
The first three stages (filtering, splitting, field extraction) are computationally intensive; instead of blindly splitting packets by byte borders, they process raw packet captures byte-by-byte, searching for specific fields and flags.
We implement them in C++ using PcapPlusPlus~\cite{pcapplusplus}, providing an order-of-magnitude speedup over pure-Python alternatives.
This investment is essential because the 4-billion-flow pretraining corpus would be infeasible to process in Python within a reasonable time budget; efficient intermediate representation structure makes the full data pipeline a one-time cost rather than a recurring bottleneck.
The pipeline supports both pretraining (unlabeled) and fine-tuning (labeled) modes, and produces Apache Arrow shards that enable memory-mapped streaming, directly integrating with HuggingFace interface and accelerating data access by workers. In addition, we split big datasets into hundreds of shards to enable parallel data access for training.

\smartparagraph{Containerized, reproducible workflow.}
The model's entire workflow (raw PCAP ingestion through preprocessing, pretraining, and fine-tuning) is containerized via Docker and orchestrated through a single Makefile.
The Docker image bundles all system-level dependencies, including PcapPlusPlus for C++ packet parsing, CUDA toolkits, and Python libraries with pinned versions.
Containerization matters because network traffic analysis tools have notoriously fragile dependency chains (specific versions of libpcap, protocol dissectors, CUDA drivers), and a user should be able to reproduce any experiment by simply providing input PCAPs and a configuration file, without any manual environment setup.

\subsection{Extensibility and Testing}

All aspects of \system, such as model architecture, protocol field selection, tokenizer behavior, training hyperparameters, and loss weights, are controlled through a unified, hierarchical configuration system compatible with HuggingFace configurations.
This makes it straightforward to extend \system to new protocols or customize the representation for a specific downstream task without modifying source code.

The codebase is organized into self-contained modules (tokenizer, embeddings, layers, attention, collator, sampler, trainer, metrics, configuration, and utilities), each with a well-defined interface.
The fine-tuning and pretraining models share a common base encoder, so downstream tasks can be added by implementing a new prediction head without modifying the core model.
The trainer extends the HuggingFace Trainer with domain-specific columns and callbacks, preserving compatibility with the broader ecosystem (e.g., Weights~\&~Biases logging, DeepSpeed integration, gradient accumulation strategies, Accelerate, etc).

We maintain a model-specific test suite covering core infrastructure components: the custom tokenizer, length-bucketed sampler, data collator with burst swapping and metadata masking, attention mask transformations, and tensor reshape operations.
These tests exercise edge cases (e.g., empty buffers, single-example batches, variable burst lengths) and verify numerical correctness using parametrized fixtures, providing confidence that modifications to the codebase or model's extensions do not introduce silent regressions.

\section{Evaluation}


In this section, we evaluate and analyze \system's performance across multiple dimensions, including intrinsic representation quality of the encoder's embeddings and downstream task performance. We also report on the overall computational efficiency of \system in \autoref{sec:eval-efficiency}.

\smartparagraph{Models and datasets.}
We compare \system (small, base, and large checkpoints) against five publicly available network foundation models that represent distinct design philosophies: ET-BERT~\cite{etbert}, a BERT-based model pretrained on encrypted payload bytes; YaTC~\cite{yatc}, a vision-transformer approach that treats traffic as images; NetMamba~\cite{netmamba}, a state-space model for network traffic; TrafficFormer~\cite{trafficformer}, a BERT-style transformer using both packet headers and payload information; and Pcap-Encoder~\cite{debunking}, a T5-based model proposed alongside a corrected evaluation protocol.
For all competitors, we use the publicly released pretrained checkpoints without modification.
The datasets used for intrinsic and downstream evaluations differ, and the relevant differences are detailed in the next two subsections. Note that for the reasons mentioned in~\autoref{sec:limitations}, we intentionally omitted an ablation study.

\subsection{Intrinsic Evaluation}
\label{sec:eval-intrinsic}
Recent works~\cite{demystifying} demonstrated that evaluating network foundation models solely through downstream fine-tuning performance conflates the quality of pretrained representations with confounding factors such as model capacity, classification head design, and optimization dynamics, making it difficult to attribute gains to the pretraining itself.
An intrinsic evaluation that directly probes the learned representations in a task-agnostic manner isolates the contributions of these confounding factors and allows for a more meaningful assessment of what the encoder has actually learned during pretraining.

Here, we adopt the Intrinsic Evaluation Framework (IEF)~\cite{demystifying}, a task-agnostic methodology that probes the representational quality of network foundation models through three complementary lenses: embedding geometry analysis quantifies how effectively models distribute representations across the latent space; metric alignment assessment measures correspondence between learned embeddings and established features derived by domain experts; and causal sensitivity testing evaluates whether models capture higher-order network context invisible in the raw traffic bytes, such as congestion control behavior and queue management policies.

In this subsection, we rely on the four datasets that were used in IEF: Crossmarket~\cite{crossmarkets}, CIC-IDS-2017~\cite{cic-ids-2017}, CAIDA~\cite{caida}, and MAWI~\cite{mawi}. These datasets cover a range of different traffic conditions, including endogenously (synthetically) and exogenously (production) generated data, and provide a comprehensive suite for intrinsic evaluation. 

\subsubsection{Embedding Geometry}


Mean cosine similarity between random pairs of flow embeddings measures the average distance between different network flows from the model's perspective: the further the distance, the more different these flows are in the embedding space of the model. For a given dataset, this average distance is expressed in terms of the anisotropy metric, where values close to 1 indicate that the model collapses flows into nearly identical representations, while lower values suggest that the model distributes information across the available dimensions and can distinguish meaningful variations in network behavior. At the same time, values close to 0 demonstrate that the model cannot find any similar flows and spread embeddings far, making clusterization and analysis impossible. We calculate mean cosine similarity between the embeddings produced for the benchmark datasets and report the results in~\autoref{tab:anisotropy}.

\begin{table}[htbp]
    \caption{Mean cosine similarity between flow embeddings. Bold shows results that differ significantly from competitors.}
    \centering
    \label{tab:anisotropy}
    \resizebox{\linewidth}{!}{
    \begin{tabular}{l|ccccc}
        Model & Crossmarket & CIC-IDS-2017 & CAIDA & MAWI & Avg. \\
        \hline
        Random        & 0.00 & 0.00 & 0.00 & 0.00 & 0.00 \\
        ET-BERT       & 0.88 & 0.74 & 0.87 & 0.85 & 0.84 \\
        YaTC          & 0.85 & 0.85 & 0.86 & 0.85 & 0.85 \\
        NetMamba      & 0.93 & 0.92 & 0.99 & 0.99 & 0.96 \\
        TrafficFormer & 0.74 & 0.68 & \textbf{0.60} & \textbf{0.54} & \textbf{0.64} \\
        Pcap-Encoder  & 0.86 & 0.77 & 0.69 & 0.84 & 0.79 \\
        \hline
        \system-small & 0.68 & 0.65 & 0.79 & 0.78 & 0.72 \\
        \system-base  & \textbf{0.63} & \textbf{0.60} & 0.73 & 0.77 & \textbf{0.68} \\
        \system-large & \textbf{0.64} & \textbf{0.55} & 0.68 & 0.76 & \textbf{0.66} \\
    \end{tabular}
    }
\end{table}

\smartparagraph{Takeaway.}
Our findings demonstrate that while YaTC and NetMamba collapse embeddings into near-identical representations, TrafficFormer, Pcap-Encoder, ET-BERT, and all versions of \system maintain substantially lower cosine similarity, indicative of a non-degenerate geometry. TrafficFormer achieves the lowest average value (0.64), but anisotropy is a diagnostic rather than a winner-take-all metric: lower cosine similarity is only of relevance if it reflects some meaningful structure, and TrafficFormer's low anisotropy is plausibly explained by its access to encrypted payload bytes. While these bytes are not semantically informative, they can be expected to add input diversity and result in richer geometric structure.

When examined through the lens of anisotropy, \system achieves the second-lowest average cosine similarity (0.66--0.72), preserves user's privacy, and outperforms TrafficFormer on the endogenous Crossmarket and CIC-IDS-2017 datasets, where \system-large achieves 0.64 and 0.55 versus 0.74 and 0.68. These findings confirm that the combination of protocol-aware tokenization (Principle~1), operational context embedding (Principle~2), and burst-flow hierarchical attention (Principle~3) helps to distribute representations more uniformly across the latent space, directly addressing the embedding collapse problem identified by IEF. We also note that \system typically exhibits higher anisotropy on the exogenous CAIDA and MAWI datasets due to them containing less expressed packet header variations.
Given that the privacy-by-construction input design (Principle~4) excludes any payload information, this is expected behavior rather than a model limitation.


\subsubsection{Metric Alignment}

In IEF, Centered Kernel Alignment (CKA)~\cite{kornblithSimilarityNeuralNetwork2019} measures the structural similarity between a model's flow embeddings and domain expert-derived features (e.g., flow duration, packet size distributions, etc.) computed by CICFlowMeter~\cite{lashkariCharacterizationTorTraffic2017}. This metric measures whether the hidden network representation from a foundation model indeed contains the statistical information that correlates with each flow and is used by network domain experts and does not simply represent the raw byte values of the packets. Higher CKA values (near 1) indicate that the model's latent space implicitly encodes well-known features, validating that it has learned generalizable network semantics rather than superficial correlations. We calculate the CKA metric for each model and report the computed values in~\autoref{tab:cka}. For comparison, we also report the CKA values for randomly generated embeddings and for oracle embeddings that directly represent the computed CICFlowMeter features.

\begin{table}[htbp]
    \caption{CKA similarity between model embeddings and CICFlowMeter features (higher is better).}
    \centering
    \label{tab:cka}
    \resizebox{\linewidth}{!}{
    \begin{tabular}{l|ccccc}
        Model & Crossmarket & CIC-IDS-2017 & CAIDA & MAWI & Avg. \\
        \hline
        Random        & 0.000 & 0.000 & 0.000 & 0.000 & 0.000 \\
        ET-BERT       & 0.012 & 0.064 & 0.033 & 0.026 & 0.047 \\
        YaTC          & 0.098 & 0.092 & 0.014 & 0.070 & 0.068 \\
        NetMamba      & 0.047 & 0.042 & 0.030 & 0.051 & 0.042 \\
        TrafficFormer & 0.138 & 0.083 & \textbf{0.053} & 0.057 & 0.082 \\
        Pcap-Encoder  & 0.103 & 0.096 & 0.036 & \textbf{0.082} & 0.079 \\
        Oracle (ideal) & 1.000 & 1.000 & 1.000 & 1.000 & 1.000 \\
        \hline
        \system-small & \textbf{0.154} & \textbf{0.149} & 0.046 & 0.059 & \textbf{0.102} \\
        \system-base  & \textbf{0.154} & \textbf{0.143} & 0.045 & 0.030 & \textbf{0.093} \\
        \system-large & \textbf{0.155} & \textbf{0.144} & 0.046 & 0.033 & \textbf{0.094} \\
    \end{tabular}
    }
\end{table}

\smartparagraph{Takeaway.}
All versions of \system achieve the highest average CKA (0.093--0.102), with the best competitors (Pcap-Encoder and TrafficFormer) scoring 0.079 and 0.082, respectively. 
The gap is most pronounced for the endogenous datasets (Crossmarket and CIC-IDS-2017), where \system achieves 0.154 and 0.149, improving over 0.138 and 0.096 from the closest competitors (TrafficFormer and Pcap-Encoder). 
These results confirm that the explicit operational context embedding (Principle~2) grounds \system's representations in the operational features that domain experts rely on. The relatively modest alignment results for the CAIDA and MAWI datasets are consistent with those datasets' skewed ratio of shorter network flows which prevents the CICFlowMeter features from exhibiting realistic variability, causing \system's CKA values to be close to those of the competitors.


\subsubsection{Exogenous Network Context}

To test whether models capture unobserved higher-order network conditions, we use the synthetically generated flows  published by IEF~\cite{demystifying} which contain controlled variations in congestion control algorithm, active queue management policy, and cross-traffic patterns. Since these flows represent the same application-level behavior and have identical payload under different network contexts, they allow us to examine whether embeddings reflect these contexts or simply memorize the encrypted data.
A model that meaningfully encodes these exogenous factors should produce embeddings that are linearly separable by context, a property that can be measured via a logistic-regression probe using an $F_1$ score.
As noted in IEF, this evaluation differs from a downstream performance evaluation in the sense that it utilizes the simplest linear probe design to only assess a linear combination of the encoder's output embeddings. In effect, this evaluation isolates the representational quality of the pretrained model and prevents conflating it with the capacity of a learned classification head.

\begin{table}[htbp]
    \caption{Exogenous context discrimination: linear probing performance using $F_1$ score (higher is better). CC, AQM, CTP, and All represent Congestion Control, Active Queue Management, Cross-Traffic Patterns classifications, and all changes combined respectively.}
    \centering
    \label{tab:context}
    \small
    \begin{tabular}{l|cccc}
        Model & CC & AQM & CTP & All \\
        \hline
        ET-BERT       & 0.41 & 0.68 & 0.43 & 0.46 \\
        YaTC          & 0.48 & 0.51 & 0.24 & 0.47 \\
        NetMamba      & 0.29 & 0.70 & 0.33 & 0.62 \\
        TrafficFormer & 0.36 & 0.55 & 0.49 & 0.57 \\
        Pcap-Encoder  & 0.00 & 0.68 & 0.17 & 0.59 \\
        \hline
        \system-small & \textbf{0.64} & \textbf{0.86} & \textbf{0.79} & \textbf{0.90} \\
        \system-base  & \textbf{0.57} & \textbf{0.84} & \textbf{0.74} & \textbf{0.93} \\
        \system-large & \textbf{0.65} & \textbf{0.90} & \textbf{0.81} & \textbf{0.95} \\
    \end{tabular}
\end{table}

\smartparagraph{Takeaway.}
All three versions of \system achieve excellent $F_1$ scores (0.90--0.95) on the combined exogenous context discrimination task (``All''). In contrast, the score of the best competitor (NetMamba) is 0.62. We observe the same pattern across all individual contexts: even \system-small scores 0.64/0.86/0.79 on CC/AQM/CTP versus the best competitors scores of 0.48/0.70/0.49 (YaTC, NetMamba, TrafficFormer).

These findings are particularly telling. Since the flows have the same payload across different contexts, a model must rely on temporal patterns (e.g., inter-packet arrival times, burst structures) to achieve a high level of discrimination. For \system, the operational context embedding (Principle~2) and burst-flow hierarchical attention mechanism (Principle~3) are the enablers: congestion control, queue management, and cross-traffic shape the \emph{temporal} structure of bursts rather than packet content, and \system's architecture is explicitly designed to capture these multi-scale patterns. This observation reinforces the benefits of the diagnostic-to-design approach advocated in this paper: the principles derived from the context discrimination findings reported in~\cite{demystifying} give rise to a model that excels exactly at the task of network context discrimination. 


\subsection{Downstream Performance}
\label{sec:eval-fine}

\begin{table*}[!t]
    \caption{Weighted $F_1$ score on flow classification downstream tasks with bold highlighting the results that are significantly higher than competitors. Underlined red scores show benchmarks that were utilized during pretraining by corresponding models. Even in this unfair setting, \system demonstrates overwhelming performance for frozen and on par for unfrozen scenarios.}
    \label{tab:finetuning}
    \centering
    \resizebox{\textwidth}{!}{%
    \begin{tabular}{l|cc|cc|cc|cc|cc}

        \multirow{2}{*}{Model} 
        & \multicolumn{2}{c|}{USTC (20 classes)} 
        & \multicolumn{2}{c|}{USTC (binary)} 
        & \multicolumn{2}{c|}{ISCX-VPN} 
        & \multicolumn{2}{c|}{CIC-IDS-2017} 
        & \multicolumn{2}{c}{Crossmarket (Acc@10)} \\ 
        & Frozen & Unfrozen 
        & Frozen & Unfrozen 
        & Frozen & Unfrozen 
        & Frozen & Unfrozen 
        & Frozen & Unfrozen \\
        \hline
        ET-BERT & 0.8533 & 0.8935 & \textbf{0.9998} & 0.9999 & \leakedresult{0.6684} & \leakedresult{0.8645} & 0.8580 & 0.8783 & 0.1857 & 0.3283 \\
        YaTC & \leakedresult{\textbf{0.9396}} & \leakedresult{0.9396} & \leakedresult{0.9974} & \leakedresult{0.9978} & \leakedresult{0.5603} & \leakedresult{0.5603} & 0.9843 & 0.9999 & 0.2077 & 0.6348 \\
        NetMamba & \leakedresult{0.4191} & \leakedresult{\textbf{0.9881}} & \leakedresult{0.6496} & \leakedresult{0.9999} & \leakedresult{0.1362} & \leakedresult{\textbf{0.9364}} & 0.9146 & 0.9998 & \leakedresult{0.1733} & \leakedresult{0.7149} \\
        TrafficFormer & \textbf{0.9555} & 0.9789 & \textbf{0.9999} & 1.0000 & \leakedresult{0.6459} & \leakedresult{0.8712} & \textbf{0.9916} & 0.9995 & 0.2322 & 0.5923 \\
        Pcap-Encoder & 0.8799 & 0.9775 & \textbf{1.0000} & 0.9999 & 0.3967 & 0.9032 & 0.9691 & 0.9883 & 0.2257 & \textbf{0.8699} \\
        \hline
        \system-small & 0.8550 & 0.9645 & \textbf{0.9999} & 1.0000 & \textbf{0.7508} & 0.8920 & \textbf{0.9928} & 0.9996 & \textbf{0.3804} & 0.4727 \\
        \system-base & 0.8908 & 0.9695 & \textbf{0.9999} & 1.0000 & \textbf{0.8053} & \textbf{0.9175} & \textbf{0.9918} & 0.9997 & \textbf{0.4330} & 0.6066 \\
        \system-large & \textbf{0.9259} & 0.9674 & \textbf{0.9999} & 1.0000 & \textbf{0.8098} & 0.9004 & \textbf{0.9938} & 0.9998 & \textbf{0.4561} & 0.6538 \\
    \end{tabular}
    }
\end{table*}

To assess the practical utility of representations, we evaluate \system on five downstream tasks spanning encrypted traffic classification, malware detection, intrusion detection, and application fingerprinting.
We compare against the same five published state-of-the-art network foundation models: ET-BERT~\cite{etbert}, YaTC~\cite{yatc}, NetMamba~\cite{netmamba}, TrafficFormer~\cite{trafficformer}, and Pcap-Encoder~\cite{debunking}. 
All models are evaluated in two regimes: \textbf{frozen}, where the pretrained encoder weights are fixed and only the model-defined head is trained on top of the encoder's output representation, and \textbf{unfrozen}, where the full model is fine-tuned end-to-end.
The frozen setting allows us to directly assess the quality of the learned representations: a high score is a strong indication that the pretrained embeddings already capture task-relevant structure, making the model usable as a feature extractor without expensive retraining.
The unfrozen setting enables us to measure the model's overall capacity when all parameters can be adapted to the target task, including the model's potential for learning a dataset's shortcuts.
Importantly, where prior work explicitly benefited from leveraging inconspicuous dataset shortcuts (see Pcap-Encoder~\cite{debunking}, \autoref{tab:resilience}, and the discussion in \S\ref{sec:background-shortcuts}), the results we report below are based on the \emph{corrected} versions of datasets that prevent learning the mentioned shortcuts.

\smartparagraph{Datasets.}
\autoref{tab:downstream-datasets} summarizes the five downstream benchmarks used in our evaluation, chosen to cover a diverse range of task types, class granularities, and traffic conditions.

\begin{table}[htbp]
    \caption{Downstream evaluation datasets.}
    \centering
    \label{tab:downstream-datasets}
    \resizebox{\linewidth}{!}{%
    \begin{tabular}{l|l|r|l}
        Dataset & Task & Classes & Metric \\
        \hline
        USTC-TFC~\cite{ustc-tfc} & App classification & 20 & $F_1$ \\
        USTC-TFC~\cite{ustc-tfc} & Malware detection & 2 & $F_1$ \\
        ISCX-VPN~\cite{DraperGil2016CharacterizationOE} & VPN traffic classif. & 17 & $F_1$ \\
        CIC-IDS-2017~\cite{cic-ids-2017} & Intrusion detection & 8 & $F_1$ \\
        Crossmarket~\cite{crossmarkets} & App fingerprinting & 210 & $Acc@10$ \\
    \end{tabular}
    }
\end{table}

The USTC-TFC dataset provides two complementary classification tasks over the same encrypted traffic: a fine-grained 20-class application identification task and a binary malware-vs-benign detection task, testing both multi-class discrimination and coarse security triage.
ISCX-VPN evaluates the ability to characterize traffic within encrypted tunnels, a setting where header information is partially obscured.
CIC-IDS-2017 targets intrusion detection across 7 attack categories (a separate 8th category is used for benign traffic), stressing the model's capacity to distinguish subtle behavioral anomalies.
Finally, Crossmarket is a large-scale application fingerprinting benchmark with 210 application classes, evaluated via $Acc@10$ (accuracy within the top-10 predictions), which allows for rigorous representation quality testing that is especially relevant for retrieval- and similarity-based tasks.

\subsubsection{Analysis}


The frozen columns in \autoref{tab:finetuning} reveal how much task-relevant structure the pretrained encoder captures \emph{before any task-specific encoder weight updates}.
Across \system's three model sizes, frozen representations are already highly discriminative: even the smallest model variant achieves weighted $F_1$ scores of $0.75-0.99$, demonstrating that a 53M-parameter frozen encoder can efficiently represent diverse network flows.

A clear scaling trend emerges in the frozen setting.
On USTC-20, the frozen $F_1$ scores increase monotonically from $0.855$ (small) to $0.891$ (base) to $0.926$ (large), a 7-point improvement driven solely by additional model capacity during pretraining.
The same pattern holds for ISCX-VPN ($0.751 \to 0.805 \to 0.810$) and Crossmarket ($0.380 \to 0.433 \to 0.456$), confirming that larger models internalize richer traffic representations.
The effect is most pronounced on more challenging tasks: Crossmarket, which requires discriminating among 210 applications via ranking, shows a relative improvement of $\sim$20\% from small to large, whereas nearly-saturated tasks like USTC binary show no headroom for further gains.

The observed frozen-to-unfrozen gap can be viewed as providing an additional diagnostic tool.
On tasks where frozen scores are already high (USTC-binary and CIC-IDS-2017), fine-tuning yields negligible additional benefit, indicating that the pretrained representations already capture the relevant decision boundary.
Conversely, tasks with a larger frozen-to-unfrozen gap, such as VPN ($0.751 \to 0.892$ for small) and Crossmarket ($0.380 \to 0.473$ for small), suggest that while the pretrained model provides a good starting point, task-specific adaptation can unlock additional performance.
Notably, this gap narrows with scale: for USTC-20 and ISCX-VPN, the gap between the small and large model variants shrinks ($0.109 \to 0.042$ and $0.141 \to 0.091$, respectively). Moreover, on Crossmarket, the large model's frozen score ($0.456$) already approaches the small model's unfrozen score ($0.473$), illustrating that scaling pretraining can partially substitute for task-specific fine-tuning. 

In addition to being the best model in the frozen setting, the unfrozen version of \system emerges as the top performer across all the benchmarks, exceeding or matching the competitors' performance in most of the cases (see~\autoref{tab:finetuning}). This property holds even for the small model variant, indicating that \system's architecture and pretraining strategy provide a practical and exceptionally strong starting point for downstream adaptation.

\smartparagraph{Test-into-train leakage.} It is worth noting that for many models, their reported performance is not only a function of architectural decisions but, due to the limited number of publicly available pretraining datasets, can also be impacted by the use of pretraining data that includes fine-tuning datasets. The use of such augmented pretraining data allows models to generate good quality fine-tuning task-specific embeddings but limits their ability to generalize to previously unseen data. The results for such models and datasets are highlighted in red in~\autoref{tab:finetuning}. We observe that most of the frozen models achieve low $F_1$ scores on the Crossmarket dataset, which was not used for pretraining by any of the models with the exception of NetMamba. \system's choice of pretraining data prevents any leakage of test data to the frozen encoder, resulting in smaller gaps between frozen and unfrozen setting than other models. Even in this disadvantageous for \system setting and despite its fully privacy-preserving design (i.e., no payload data), \system achieves the best performance in the frozen setting and is comparable with the best competitors in the unfrozen setting. 
\section{Limitations}
\label{sec:limitations}


\smartparagraph{Absence of Full Ablation Study.}
\system incorporates multiple interacting design decisions: protocol-aware tokenization, operational context embedding, burst-flow hierarchical attention, four pretraining objectives, and others.
A comprehensive ablation study would require pretraining each variant of \system from scratch for every combination of removed substituted model components, consuming thousands of GPU-hours on up to 128 A100 GPUs, rendering the resulting cost prohibitive. Large language model efforts (e.g., LLaMA~\cite{llama}, ModernBERT~\cite{modernbert}) face a similar problem and omit ablation studies over their design space for the same reasons.
Instead, we justify individual design choices through (i) prior evidence in the literature (e.g., RoPE, GeGLU, Flash Attention have been extensively validated in NLP), (ii) domain-specific motivation grounded in the structure of network data (e.g., hierarchical attention mirrors the burst/flow hierarchy), and (iii) intrinsic and downstream evaluation of the final trained models across multiple tasks and datasets.
We acknowledge that isolating the marginal contribution of each component remains an open question, and we encourage future efforts that are less encumbered by tight compute budgets to explore this topic.

\smartparagraph{Encoder-Only Architecture.}
\system is built on an encoder-only transformer, which makes it well suited to discriminative tasks such as classification, anomaly detection, and traffic characterization. However, the current design precludes the use of \system for generative capabilities, e.g. synthetic traffic generation, packet-level forecasting, or open-ended reasoning about network traffic traces.
Extending the model to an encoder-decoder or decoder-only design to support such generative use cases is a promising direction for future work.

\smartparagraph{Pretraining data limitations.}
The single-campus pretraining data is geographically concentrated and U.S.-specific, potentially limiting the applicability of the pretrained models to different environments, such as enterprise datacenter networks, IoT ecosystems, or countries with significantly different backbone traffic. To mitigate this limitation, we release the full source code so as to facilitate pretraining on user-specific data and encourage the reproducibility of our approach.

\smartparagraph{Reliance on evaluation benchmarks.}
Pcap-Encoder~\cite{debunking} and other works~\cite{9947235, demystifying, trustee} demonstrated that many existing evaluation benchmarks can have shortcut-susceptible label distributions, are based on problematic evaluation designs, or can be biased towards measuring specific model qualities. Given that our design effort is in large parts inspired by the findings reported in these prior works, we acknowledge a potential bias in our choice of relevant design principles. To keep an eye on such a potential bias and prevent it from affecting our approach, we examine and validate our design decisions from multiple perspectives (e.g., intrinsic evaluation, studying frozen and unfrozen model settings, and efficiency).
\section{Conclusion}


In this paper, we present \system, a diagnostic-informed, production-ready network foundation model.
Building on the NFM failure analysis conducted by preliminary works, we translate these findings into four network-native design principles and realize them in an efficient, privacy-preserving, system-driven approach, implementing multiple novel architectural components and training techniques.
Across intrinsic evaluation, frozen, and unfrozen downstream analysis, \system demonstrates stronger reusable representations via lower anisotropy, higher correlation with domain-expert features, better sensitivity to exogenous network context, robustness to shortcuts, and top performance across all five downstream tasks. 
We pair this model with a full release of code, pretrained checkpoints, the pretraining dataset, and an end-to-end pipeline from raw PCAPs to inference so that the community can reproduce, extend, and deploy the system in practice. 
By this, we aim to provide a foundation for future application of diagnosis-to-design methodology and motivate the community to drive the next generation of architectural improvements as new diagnostic tools and public datasets emerge.

\section*{Availability}
\label{sec:availability}

We make \system publicly available as a reusable research artifact.
The release includes the pretrained checkpoints for all three model sizes (small at \url{https://huggingface.co/snlucsb/netFound-small}, base at \url{https://huggingface.co/snlucsb/netFound-base}, and large at \url{https://huggingface.co/snlucsb/netFound-large}) on HuggingFace, as well as the source code for system at \url{https://github.com/SNL-UCSB/netFound} and all claims from this paper at \url{https://github.com/maybe-hello-world/netfound-paper}. In addition, we also \textit{publish the full pretraining dataset}, containing more than 4 billion network flows in Apache Arrow format, suitable for pretraining of \system or any future model utilizing a similar tokenizer. The full dataset download instructions are available in the \system GitHub repository at \url{https://github.com/SNL-UCSB/netFound}, and a separate sampler subset of the data is available at \url{https://zenodo.org/records/19863446}. Together, these artifacts provide an end-to-end, reproducible pipeline for training and applying \system to various network analysis tasks and problems.
\section*{Acknowledgments}

This work was supported by the National Science Foundation (CAREER Award No. 2443777 and CNS Award No. 2323229), research gifts from Cisco and Google, and a subcontract from Lawrence Berkeley National Laboratory's Energy Sciences Network (ESnet). This research used resources of the National Energy Research Scientific Computing Center (NERSC), a DOE Office of Science User Facility supported by the Office of Science of the U.S. Department of Energy under Contract No. DE-AC02-05CH11231 using NERSC award NERSC DDR-ERCAP0029768.

\bibliographystyle{plain}
\bibliography{ref}

\newpage
\appendix

\section{Cost \& Efficiency}
\label{sec:eval-efficiency}

\begin{strip}
    \centering
    \captionof{table}{Unfrozen model performance on a single GPU A100 80Gb with BF16 enabled.}
    \label{tab:performance}
    \begin{tabular}{l|r|r|r|r|r|r|r|r}
        \multirow{2}{*}{Model} 
        & \multirow{2}{*}{Params} 
        & \multirow{2}{*}{Batch} 
        & \multicolumn{3}{c|}{Training}
        & \multicolumn{3}{c}{Inference} \\ 
        & & & Mem (MB) & Latency (ms) & flows/sec & Mem (MB) & Latency (ms) & flows/sec \\
        \hline
        small & \multirow{3}{*}{53M}  & 1  & 1,549  & $34.9 \pm 1.2$   & $29 \pm 1$ & 803   & $11.6 \pm 0.3$ & $86 \pm 2$ \\
        small &                       & 8  & 2,641  & $37.7 \pm 0.7$   & $212 \pm 4$ & 967   & $15.2 \pm 0.9$ & $528 \pm 32$ \\
        small &                       & 32 & 5,951  & $81.9 \pm 2.6$   & $390 \pm 13$ & 1,563 & $38.5 \pm 0.7$ & $832 \pm 16$ \\
        base & \multirow{3}{*}{225M}  & 1  & 3,899  & $89.3 \pm 1.6$   & $11 \pm 1$ & 1,527 & $29.8 \pm 0.4$ & $33 \pm 0.5$ \\
        base &                        & 8  & 8,189  & $106.4 \pm 7.4$  & $75 \pm 6$ & 1,933 & $40.0 \pm 0.8$ & $200 \pm 4$ \\
        base &                        & 32 & 22,103 & $303.0 \pm 3.6$  & $106 \pm 1$ & 2,871 & $116.3 \pm 2.6$ & $275 \pm 6$ \\
        large & \multirow{3}{*}{663M} & 1  & 12,591 & $185.2 \pm 9.7$  & $5 \pm 1$ & 4,521 & $56.3 \pm 0.8$ & $18 \pm 1$ \\
        large &                       & 8  & 28,353 & $277.8 \pm 33.9$ & $29 \pm 2$ & 5,225 & $100.0 \pm 9.1$ & $80 \pm 8$ \\
        large &                       & 32 & 77,307 & $952.4 \pm 8.9$  & $34 \pm 1$ & 6,717 & $337.8 \pm 1.1$ & $96 \pm 1$ \\
    \end{tabular}
\end{strip}


To support deployment decisions of \system, we report the model's computational profile across different sizes.
These measurements validate the engineering investments described in~\autoref{sec:implementation} and enable practitioners to select the appropriate model variant for their deployment constraints.

\smartparagraph{Training Cost.}
We utilize a dynamic pretraining strategy, beginning with 128xA100 40Gb GPUs for the initial pretraining stage and reducing the number of GPUs to 80 for subsequent stages together with lowering learning rate and changing objectives weights. In total, the entire pretraining process for all three model variants consumed $\sim$5,000 GPU-hours, with small, base, and large models pretraining for approximately 45K, 75K, and 35K steps respectively, and took $\sim$12-36hrs for each checkpoint.
To provide a comparison baseline, NVIDIA estimates pretraining of a single 220M T5 base checkpoint (which is used in Pcap-Encoder) from scratch as $\sim$3000 GPU hours~\cite{nvidia-nemo} and 1T of tokens using internal highly optimized training pods setup. Our lower pretraining cost, together with intrinsic and extrinsic evaluation results, highlights the benefits of network-optimized architecture.


\smartparagraph{Throughput and Memory Footprint.}
\autoref{tab:performance} shows peak consumed GPU memory (in MB), latency per batch (in ms), and throughput (flows/sec) for both training and inference across different model and batch sizes. The three model sizes offer a clear throughput--accuracy trade-off, as low as 11ms per flow, enabling deployment on hardware ranging from edge devices (small) to GPU clusters (large). Compared with Pcap-Encoder, which uses 2,989 MB GPU memory and 230ms for a single inference step with a batch size of 32 on the same hardware, our base model achieves 2x faster inference with the same number of parameters and memory taken.




\smartparagraph{Preprocessing Throughput.}
The C++-based preprocessing pipeline described in \S\ref{sec:implementation} processes 2000 flows/sec during filtering, 216 flows/sec during splitting, and 970 flows/sec during fields extraction on 128 cores, taking in total 12hrs to process a single dataset chunk of 7mln flows. Tokenization and Apache Arrow shard assembly add 4 hours walltime (500 flows/sec in Python). As a result, the full 4-billion-flow pretraining corpus is estimated to take $\sim$7000 CPU hours on a single 128-core machine. Implementing the same preprocessing in pure Python is estimated to take x10-x100 more walltime for the full dataset. To reduce efforts for preprocessing a dataset of such scale, we also publish the full preprocessed pretraining dataset (see \hyperref[sec:availability]{Availability} for details).

\section{Linear transformations in embedding space}

\begin{table}[htbp]
    \centering
    \caption{Cosine similarity and $L_1$ distance from the resulting embedding $E_R$ to embeddings of network
    data with various high-level context}
    \label{tab:synth_math}
    \resizebox{\linewidth}{!}{%
        \begin{tabular}{lcccccc}
        & \multicolumn{2}{c}{\system-small} & \multicolumn{2}{c}{\system-base} & \multicolumn{2}{c}{\system-large} \\
        & \textit{cos} & $L_1$ & \textit{cos} & $L_1$ & \textit{cos} & $L_1$ \\
        \hline
        $E_R$ to $E_{Base}$  & 0.9803 & 78.5 & 0.9561 & 105.6 & 0.9722 & 136.9 \\
        $E_R$ to $E_{CC}$    & 0.9820 & 74.7 & 0.9554 & 105.6 & 0.9731 & 135.8 \\
        $E_R$ to $E_{AQM}$   & 0.9778 & 82.8 & 0.9492 & 113.4 & 0.9684 & 148.2 \\
        $E_R$ to $E_{Cross}$ & 0.9691 & 99.3 & 0.9302 & 134.3 & 0.9600 & 168.5 \\
        \hline
        $E_R$ to $E_{All}$   & \textbf{0.9880} & \textbf{65.7} &\textbf{ 0.9569} & \textbf{102.3} & \textbf{0.9769} & \textbf{132.2} \\
        \end{tabular}
    }
\end{table}

Linear transformations in the embedding space are designed to demonstrate whether it is possible to generate embeddings of network traffic with given parameters without access to the traffic itself by using transformation vectors applied to different embeddings. The most popular example of this scenario was presented by Ethayarajh et al.~\cite{ethayarajh-etal-2019-towards}, which demonstrated the transformation \textit{king + (woman - man) = queen} in the word embedding space. Similar to the word embeddings, we explore whether \system's embeddings uphold similar properties in the network embedding space.

We use the synthetic networking dataset provided by IEF~\cite{demystifying} (the same as used in \autoref{tab:context}) which contains a single application traffic under different Congestion Control algorithms, AQM policies, and cross-traffic patterns. We select a certain combination of these parameters (FIFO, BBR, Cross-traffic \#50) and infer \system to generate embeddings ($E_{Base}$) for this traffic, representing a base starting point for modifications. Similarly, we extract embeddings $E_{CC}$, $E_{AQM}$, and $E_{Cross}$, which differ from the traffic for $E_{Base}$ by only using different Congestion Control, AQM policy, and Cross-traffic correspondingly.

Similar to IEF (Appendix H), we calculate embedding $E_R$ defined as $E_R = E_{CC} + E_{AQM} + E_{Cross} - 2* E_{Base}$. This embedding represents $E_{Base}$ with all three components (CC, AQM, Cross) changed to a different option. Note that embedding $E_R$ was produced without running inference on the corresponding network traffic, but only using transformations in the embedding space.

We measure cosine similarity and $L_1$ distance between the resulting embedding $E_R$ and all embeddings $E_{Base}$, $E_{CC}$, $E_{AQM}$, and $E_{Cross}$. We also infer embedding $E_{All}$ from the traffic with all three components swapped (CC, AQM, Cross-traffic) and measure distance between $E_R$ and $E_{All}$. As $E_R$ represents the transformation to obtain $E_{All}$, the cosine similarity between $E_R$ and $E_{All}$ should be the highest among all comparisons, and $L_1$ distance should be the lowest, if \system's embeddings correctly react to such transformations.

\autoref{tab:synth_math} demonstrates cosine similarity results and $L_1$ distance between $E_R$ and all other embeddings. We observe that for all three model checkpoints, \system demonstrates the highest cosine similarity and lowest distance between $E_R$ and $E_{All}$, confirming that its embeddings correctly react to the expected transformations. This confirms that it is theoretically possible to generate embeddings of network traffic in \system's embedding space using linear transformations, without access to the original traffic itself, which opens possibilities for guided network traffic generation in the future.

\end{document}